\documentclass[12pt,journal,draftcls,letterpaper,onecolumn]{IEEEtran} 

\usepackage{cite}
\usepackage{graphicx}
\usepackage{epsfig}
\usepackage{amsmath}
\usepackage{amsfonts} 
\usepackage{subfigure}
\usepackage{psfrag}
\usepackage{rotating}
\usepackage{setspace}
\usepackage{latexsym}
\usepackage{amsthm}
\usepackage{amssymb}
\usepackage{amsmath}
\usepackage{fancyvrb}
\usepackage{textcomp}
\usepackage{url}
\usepackage{ifdraft}
\usepackage{url}
\usepackage{hyperref}
\usepackage{booktabs}
\usepackage{graphicx}
\usepackage{epstopdf} 
\usepackage[latin1]{inputenc}
\hypersetup{
    bookmarks=true,         
    unicode=false,          
    pdftoolbar=true,        
    pdfmenubar=true,        
    pdffitwindow=false,     
    pdfstartview={FitH},    
    pdftitle={Performance of Amplify-and-Forward Multihop Transmission over Relay Clusters with Routing Strategies}, 
    pdfauthor={Dr. Ferkan Yilmaz, Fahd Khan and Dr. Mohamed-Slim Alouini},    
    pdfsubject={Multihop Transmission},   
    pdfcreator={Creator},   
    pdfproducer={latex.exe + dvips.exe + ps2pdf.exe}, 
    pdfkeywords={Symbol error probability, average capacity, multihop cluster network.}, 
    pdfnewwindow=true,      
    colorlinks=false,       
    linkcolor=red,          
    citecolor=green,        
    filecolor=magenta,      
    urlcolor=cyan           
} 

\theoremstyle{plain}

\theoremstyle{plain}

\theoremstyle{plain}

\theoremstyle{plain}

\newcommand{\comment}[1]{}

\DefineVerbatimEnvironment{code}{Verbatim}{fontsize=\small}
\newcommand{\figref}[1]{Fig.~\protect\ref{#1}}

\def\suscript(#1,#2,#3){{#1}^{#2}_{#3}}

\newcommand{\gammasub}[1]{{\suscript(\gamma,{},{#1})}}

\newcommand{\fracparams}[2]{\genfrac{}{}{0pt}{}{{#1}}{{#2}}}

\newcommand{\FoxY}[6][right]{
	\ifthenelse{\equal{#1}{right}}{\suscript(\rm{Y},{#2},{#3}){\left[{#4}\left|\fracparams{#5}{#6}\right.\right]}}{ 
		\ifthenelse{\equal{#1}{left}}{\suscript(\rm{Y},{#2},{#3}){\left[\left.{#4}\right|\fracparams{#5}{#6}\right]}}{
			\suscript(\rm{Y},{#2},{#3}){\left[{#4}\left|\fracparams{#5}{#6}\right.\right]} 
		} 
	}
}

\newcommand{\UIFoxH}[6][right]{
	\ifthenelse{\equal{#1}{right}}{\suscript(\mathcal{H},{#2},{#3}){\left[{#4}\left|\fracparams{#5}{#6}\right.\right]}}{ 
		\ifthenelse{\equal{#1}{left}}{\suscript(\mathcal{H},{#2},{#3}){\left[\left.{#4}\right|\fracparams{#5}{#6}\right]}}{
			\suscript(\mathcal{H},{#2},{#3}){\left[{#4}\left|\fracparams{#5}{#6}\right.\right]} 
		} 
	}
}
\newcommand{\ChaudhryZubeirIFoxH}[7][right]{
	\ifthenelse{\equal{#1}{right}}{\suscript(\rm{\Gamma},{#2},{#3}){\left[({#4},{#5})\left|\fracparams{#6}{#7}\right.\right]}}{ 
		\ifthenelse{\equal{#1}{left}}{\suscript(\rm{\Gamma},{#2},{#3}){\left[\left.({#4},{#5})\right|\fracparams{#6}{#7}\right]}}{
			\suscript(\rm{\Gamma},{#2},{#3}){\left[({#4},{#5})\left|\fracparams{#6}{#7}\right.\right]} 
		} 
	}
}

\newcommand{\FoxH}[6][right]{
	\ifthenelse{\equal{#1}{right}}{\suscript(\rm{H},{#2},{#3}){\left[{#4}\left|\fracparams{#5}{#6}\right.\right]}}{ 
		\ifthenelse{\equal{#1}{left}}{\suscript(\rm{H},{#2},{#3}){\left[\left.{#4}\right|\fracparams{#5}{#6}\right]}}{
			\suscript(\rm{H},{#2},{#3}){\left[{#4}\left|\fracparams{#5}{#6}\right.\right]} 
		} 
	}
}

\newcommand{\MeijerG}[6][right]{
	\ifthenelse{\equal{#1}{right}}{\suscript(\rm{G},{#2},{#3}){\left[{#4}\left|\fracparams{#5}{#6}\right.\right]}}{ 
		\ifthenelse{\equal{#1}{left}}{\suscript(\rm{G},{#2},{#3}){\left[\left.{#4}\right|\fracparams{#5}{#6}\right]}}{
			\suscript(\rm{G},{#2},{#3}){\left[{#4}\left|\fracparams{#5}{#6}\right.\right]} 
		} 
	}
}

\newcommand{\BesselK}[2][0]{
	\suscript({K},{},{#1})\left({#2}\right)
}

\newcommand{\Binomial}[2]{
	\setlength\arraycolsep{0.0pt}
	\left(\begin{array}{c}{#1}\\{#2}\end {array}\right)
}
\newcommand{\RealPart}[1]{
	\Re\left\{{#1}\right\}
}

\newcommand{\FourierTransform}[4][norm]{
	\ifthenelse{\equal{#1}{norm}}{
	{\mathcal{F}_{#2}\left\{{#3}\right\}\left({#4}\right)}
	}{ 
		\ifthenelse{\equal{#1}{conj}}{
		{\mathcal{F}_{#2}^{*}\left\{{#3}\right\}\left({#4}\right)}
		}{
		{\mathcal{F}_{#2}\left\{{#3}\right\}\left({#4}\right)} 
		} 
	}
}

\newcommand{\InvLaplaceTransform}[3]{
	{\mathcal{L}_{#1}^{-1}\left\{{#2}\right\}\left({#3}\right)}
}

\newcommand{\mathsym}[1]{{}}

\newcommand{\gammatilde}{\widetilde{\gamma}}

\newcommand{\manualtimesbreak}{\times\\}

\begin{document}

\title{Performance of Amplify-and-Forward  \\ Multihop Transmission over Relay Clusters \\ with Different Routing Strategies}
\author{
	\authorblockN{Ferkan Yilmaz, Fahd Ahmed Khan, and Mohamed-Slim Alouini}\\
	\vspace{2mm}
	\authorblockA{Electrical Engineering Program, King Abdullah University of Science and Technology (KAUST), \\
								Thuwal, Makkah Province, Saudi Arabia. \\
								Email(s): \{ferkan.yilmaz, fahd.khan, slim.alouini\}@kaust.edu.sa
	}
}
\maketitle
\pagestyle{empty}
\thispagestyle{empty}

\begin{abstract}
We Consider a multihop relay network in which two terminals are communicating with each other via a number of cluster of relays. Performance of such networks
depends on the routing protocols employed. In this paper, we find the expressions for the average symbol error probability (ASEP) performance of
amplify-and-forward (AF) multihop transmission for the simplest routing protocol in which the relay transmits using the channel having the best symbol to noise
ratio (SNR). The ASEP performance of a better protocol proposed in \cite{BibBoGui2009} known as the adhoc protocol is also analyzed. The derived expressions
for the performance are a convenient tool to analyze the performance of AF multihop transmission over relay clusters. Monte-Carlo simulations verify the correctness of the proposed
formulation and are in agreement with analytical results.  Furthermore, we propose new generalized protocols termed as last-$n$-hop selection protocol, the dual
path protocol, the forward-backward last-$n$-hop selection protocol, and the forward-backward dual path protocol, to get improved ASEP performances. The ASEP
performance of these proposed schemes is analysed by computer simulations. It is shown that close to optimal performance can be achieved by using the
last-$n$-hop selection protocol and its forward-backward variant. The complexity of the protocols is also studied. \vskip -10pt
\end{abstract}
\IEEEpeerreviewmaketitle

\begin{figure*}[t]
\begin{center} 
\includegraphics[width=0.85\columnwidth, keepaspectratio=true]{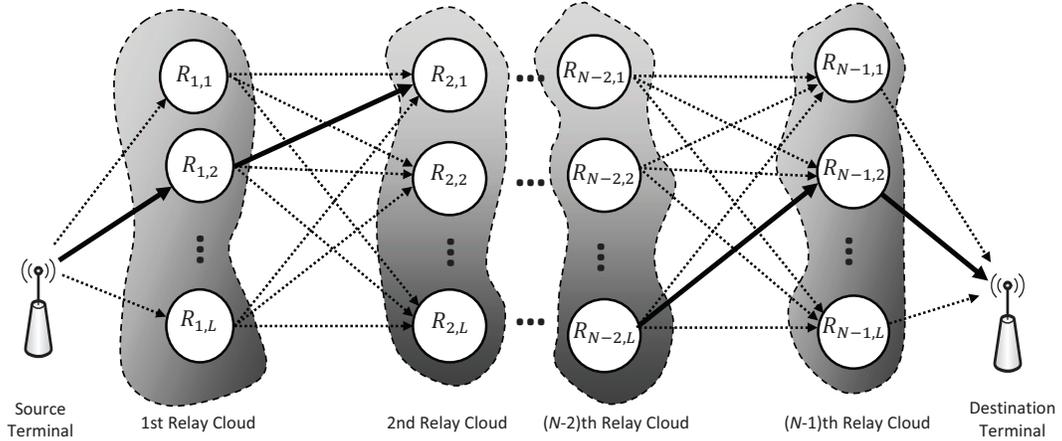}
\caption{Multihop transmission over $N-1$ relay clusters, each of which has $L$ relays.}
\label{Fig:SystemModel}
\end{center}
\vskip -35pt  
\end{figure*}

\section{Introduction}
In recent years, relaying technology has gained a lot of research interest such that relays play an increasingly important role to achieve extended coverage
not only with high data-rate transmission and the decrement of fading/shadowing effects but also without the need to employ large power at the transmitter or
relay terminals \cite{BibPabst2004}. In future wireless communication networks it is expected that the network components such as terminals will have relaying
capabilities in order to help in increasing the coverage area and to cooperate with each other by creating virtual antenna arrays, analogous to a distributed
multiple input multiple output system, to achieve improved performances \cite{BibErkip2003,BibLaneman2004,BibZheng2003}.


Information can be transmitted from the source to the destination via multiple relays as shown in \figref{Fig:SystemModel}. Routing of data through such a wireless
multihop relay network is a problem which has been extensively studied \cite{BibDraves2004Routing,BibDraves2004Comparison,BibAkkaya2005,BibTsai2006}. However
these works ignore the fading characteristic of the wireless channel and thus it is not necessary that the nearest node has the best channel. Finding the best
path by considering the signal-to-noise ratio (SNR) of the channel has been studied recently in \cite{BibZhang2008, BibBoGui2009, BibZhou2009}. The simplest
suboptimal technique to route data through the network in \figref{Fig:SystemModel} is that at each hop the relay transmits using the channel with the highest
instantaneous SNR. The performance of the system using this protocol is limited by the signal-to-noise ratio (SNR) of the channel at the last hop. This
bottle neck can be avoided by selecting the routing path for the last two hops together. This protocol is called the adhoc protocol and was proposed in
\cite{BibBoGui2009}. We will denote the former protocol as AP-1 and the adhoc protocol as AP-2.

In this paper, we derive the expressions for the average symbol error probability (ASEP) performance of the multihop network with amplify-and-forward (AF)
relays employing the AP-1 protocol and the AP-2 Protocol. The performance of the AP-2 protocol was analysed in \cite{BibBoGui2009} as well but for
a network with deocode-and-forward (DF) relays. Several papers have been published which derive the error probability performance of AF systems
\cite{BibHasna2004,BibAnghel2002,BibYilmazKucur2008,BibFerkan2010,BibDiRenzo2009A,BibDiRenzo2009B,BibSong2010}. The ASEP performance of a double hop system has
been derived in \cite{BibHasna2004,BibAnghel2002}. A comprehensive framework for the analysis of cooperative double-hop wireless systems over generalized fading
channels has been developed in \cite{BibDiRenzo2009B}. The SEP performance of multihop systems has been derived in \cite{BibYilmazKucur2008,BibFarhadi2009}.
The framework to find the exact ASEP performance for the multihop system has been presented in \cite{BibFerkan2010,BibDiRenzo2009A}. Using similar methods, the
expressions of ASEP performance for the multihop relay network are derived.

Better routing protocols can be designed by considering a cross layer routing approach \cite{BibZhang2008, BibBoGui2009, BibZhou2009}. In this paper, we propose
new routing protocols which also base the routing decisions on physical layer measurement, i.e. the instantaneous SNR of the channel. We propose a variant
of the AP-2 protocol to further improve the network performance. In the proposed protocol, denoted by AP-$n$, the routing path through the the last $n$ hops is
selected such that it has the highest end-to-end SNR. We also propose a dual path protocol in which two paths are found through the multihop relay network and
one of the paths is selected based on the best end-to-end SNR. Furthermore, usually the best path is selected from the source to destination (a forward scan).
A backward scan can also be carried out to find the best path through a network. By choosing the best route from the forward and the backward scans, a
better performance can be achieved. These protocols are termed the forward-backward routing (FBR) protocols.

For the systems using the AP-$n$ protocol (with $n>2$) or the forward-backward routing (FBR) protocols the analysis is difficult due to the correlation which
exists within the paths from source to destination. As such we determine the ASEP performance  of these protocols by computer simulations. We discuss briefly,
the implementation issues associated with the proposed protocols. Furthermore, we also give the complexity analysis of all these protocols in terms of the
number of channel state information (CSI) required and the number of comparisons required to find the path through the network.

The organization of the paper is as follows. The system model is explained in section II. The expressions for the ASEP for the AP-1 and the AP-2 protocol are
derived in section III and the performance of these protocols is illustrated by numerical results and computer simulations. The new routing protocols are
proposed in Section IV. Furthermore, the ASEP performance of the new routing protocols is analysed by numerical simulations and the implementation issues
associated with the routing protocols is also presented in Section IV. While the complexity comparison of all the routing protocols is discussed in Section V,
the performance comparison of the routing protocols is discussed in Section VI. Finally, the main results are summarized in the concluding Section VII.

\section{System Model}
We consider a $N$-hop relay network with $N-1$ clusters, a source node and a destination node. Each cluster is comprised of $L$ relays as shown in
\figref{Fig:SystemModel}. In this paper, we assume that all the relays are AF relays. In each cluster the relays are assumed to be very close with respect to the
distance between the adjacent clusters. The transmission is based on forwarding the signal to the destination via nodes in the next cluster. Time division
multiplexing is assumed and relays have the half duplex constraint. At any time only one node transmits and only one destination node (from the adjacent
cluster) listens. The channels between each relay at hop $k$ are assumed to be independent and identically distributed with Rayleigh fading and average power
$\Omega_k$, where $k={1,2,\ldots N-1}$. Note that it is not necessary that $\Omega_k=\Omega_i$ for $k \neq i$ i.e. the average power may vary from hop to hop.

\section{Average Symbol Error Probability Analysis}

In this section, we derive the expressions for average symbol error probability of the AP-$1$ system and the AP-$2$ system. Based on a moment generating
function (MGF)-approach, Yilmaz {\itshape et. al} derived the exact analytical (single integral) expressions for the evaluation of the
average error probability of wireless communication systems, for any kind of M-ary modulation schemes and for a variety number of hops, in terms of the MGF of the reciprocal of
instantaneous hop SNRs \cite{BibFerkan2010}. The end-to-end ASEP performance for the multihop transmission over multihop relay clusters can be obtained using
\begin{equation}\label{Eq:EqASEP}
P_{ASEP}=-\int_{0}^{\infty}\mathcal{Z}(u)\left[\frac{\partial}{\partial{u}}\mathcal{M}_{\gammatilde_{end}}(u)\right]du,
\end{equation}
where $\mathcal{M}_{\gammatilde_{end}}(u)$ denotes the MGF of $\widetilde{\gamma}_{end}$ and $\mathcal{Z}(u)$ is the auxiliary function defined as
\begin{equation}\label{Eq:EqInvLap}
\mathcal{Z}(u)=\InvLaplaceTransform{s}{\frac{1}{s}P_{SEP}\left(\frac{1}{s}\right)}{u},
\end{equation}
where $\InvLaplaceTransform{s}{F(s)}{u}$ is the inverse Laplace transform operator with respect to $\RealPart{s}\in\mathbb{R}^{+}$ and
$P_{SEP}\left(\gamma_{end}\right)$ is the instantaneous symbol error probability \cite{BibFerkan2010}.

\subsection{AP-1 Protocol}

For the first $N-1$ hops, there are $L$ possible channels to route the data and $\gammasub{k,{\ell}}$ denote the instantaneous SNR of
the $\ell$th possible channel at the $k$th hop, where $\ell={1,2,\ldots L}$ and $k={1,2,\ldots N-1}$. However, at the $N$th hop there is only one possible
channel to route the data and the SNR of the last hop is denoted by $\gammasub{\mathcal{D}}$. For the first $N-1$ hops, the channel with the highest
instantaneous SNR is chosen for transmission. Thus, $\gammasub{k}=\max_{\ell}\{\gammasub{k,{\ell}}\}$ denotes the SNR of channel at the $k$th hop. The
amplifying gain at the relay is chosen such that it inverts the effect of the channel of the previous hop and thus, the end-to-end SNR of the calculated route
is the harmonic mean of the SNR at each hop and is expressed as \cite{BibAlouini2003}
\begin{equation}\label{Eq:EqEndSNR}
\gammasub{end}=\frac{1}{\displaystyle\frac{1}{\gammasub{1}}+\frac{1}{\gammasub{2}}+\ldots+\frac{1}{\gammasub{N-1}}+\frac{1}{\gammasub{\mathcal{D}}}}.
\end{equation}

Using the shorthand head-script $\sim$ notation denoting that $\widetilde{\gamma}$ is the reciprocal random variable (RV) of a non-negative RV $\gamma$, we can
represent the instantaneous end-to-end SNR $\gamma_{end}$ in terms of the reciprocal of the sum of reciprocal RVs as follows   
\begin{equation}\label{Eq:EqEndSNRInReciprocality}
\widetilde{\gamma}_{end}=\sum_{k=1}^{N-1}\widetilde{\gamma}_{k}+\widetilde{\gamma}_\mathcal{D},
\end{equation}
whose probability density function (PDF), cumulative density function (CDF) and MGF are easier than those of \eqref{Eq:EqEndSNR}.

As the channels are assumed to be independent, the MGF $\mathcal{M}_{\gammatilde_{end}}(s)$ can be expressed as
\begin{equation}\label{Eq:EqEndToEndMGF}
\mathcal{M}_{\gammatilde_{end}}(s)=\mathcal{M}_{\mathcal{C}}(s)\times\mathcal{M}_{\mathcal{D}}(s),
\end{equation}
where $\mathcal{M}_{\mathcal{C}}(s)=\prod_{m=1}^{N-1}\mathcal{M}_{m}(u)$ is the MGF of the first term of
\eqref{Eq:EqEndSNRInReciprocality} (i.e., the MGF of $N-1$ terms in the denominator of \eqref{Eq:EqEndSNR}) and can be calculated from the product of the MGFs
of reciprocal hop SNRs and $\mathcal{M}_{\mathcal{D}}$ denotes the MGF of $\widetilde{\gamma}_\mathcal{D}$.

For $k \in\{1,2,\ldots,N-1\}$, the MGF of $\widetilde{\gamma}_{k}$ can be obtained using $P_{\gammasub{k}}$, i.e. the CDF of $\gammasub{k}$, as

\begin{equation}\label{Eq:EqHopMGF}
\mathcal{M}_{\widetilde{\gamma}_{k}}(s)= \int_{0}^{\infty} \exp \left(-\frac{s}{\gammasub{k}} \right)
\frac{\partial}{\partial{\gammasub{k}}} {P_{\gammasub{k}}\left( \gammasub{k} \right)} d\gammasub{k},
\end{equation}
where $P_{\gammasub{k}}$ can be easily obtained as the product of CDFs of the instantaneous SNRs of the possible channels at the $k$th hop. 
\begin{equation}\label{Eq:EqHopCDF}
P_{\gammasub{k}}\left(r\right)=P_{\gammasub{k,1}}\left(r\right)\times{P}_{\gammasub{k,2}}\left(r\right)\times\ldots\times{P}_{\gammasub{k,L}}\left(r\right).
\end{equation}

For the case of independent and identically distributed Rayleigh RVs, we obtain the MGF of $\widetilde{\gamma}_{k}=\max_{\ell}\{\gammasub{k,{\ell}}\}$ as

\begin{equation}\label{Eq:EqMGFOfOneHop}
\mathcal{M}_{\widetilde{\gamma}_{k}}(s)={2}\sum_{\ell=1}^{L}{\left(-1\right)}^{\ell+1}\binom{L}{\ell}\sqrt{\frac{\ell{s}}{\Omega_{k}}}\BesselK[1]{2\sqrt{\frac{\ell{s}}{\Omega_{k}}}},
\end{equation}
where $\BesselK[n]{\cdot}$ is the $n$-order modified Bessel function of the second kind defined in \cite[Section 6.19.14]{BibZwillingerBook}. Then, referring to
\eqref{Eq:EqEndToEndMGF}, we obtain the MGF $\mathcal{M}_{\mathcal{C}}(s)=\prod_{k=1}^{N-1}\mathcal{M}_{\widetilde{\gamma}_{k}}\left(s\right)$ as 

\begin{equation}\label{Eq:EqMcMGF}
\mathcal{M}_{\mathcal{C}}\left(s\right)={2}^{N-1}\prod_{k=1}^{N-1}\sum_{\ell=1}^{L}{(-1)^{\ell+1}}\Binomial{L}{\ell}\sqrt{\frac{\ell{s}}{\Omega_{k}}}\BesselK[1]{2\sqrt{\frac{\ell{s}}{\Omega_{k}}}},
\end{equation}
The MGF of $\widetilde{\gamma}_\mathcal{D}$ is 

\begin{equation}\label{Eq:EqMdMGF}
\mathcal{M}_{\mathcal{D}}(s)=2\sqrt{\frac{s}{\Omega_{\mathcal{D}}}}\BesselK[1]{2\sqrt{\frac{s}{\Omega_{\mathcal{D}}}}},
\end{equation}
where $\Omega_{\mathcal{D}}$ is the average power of the last hop \cite[Eq.(11)]{BibAlouini2003}. 

Finally, substituting $\mathcal{M}_{\mathcal{C}}(s)$ and $\mathcal{M}_{\mathcal{D}}(s)$ into \eqref{Eq:EqASEP}, we obtain the exact ASEP of multihop
transmission using AP-1 protocol and is shown in \eqref{Eq:EqASEPOverRaylayFadingChannels} \comment{\stepcounter{equation}}at the top of next page. The
auxiliary function $\mathcal{Z}\left(u\right)$ can be readily obtained for any kind of modulation schemes by using the inverse Laplace transform of the
instantaneous SNR $P_{SEP}\left(\gamma_{end}\right)$ \cite{BibFerkan2010}. It should be pointed out that the analytical performance presentation of
\eqref{Eq:EqASEPOverRaylayFadingChannels} is exact and its evaluation involves a single integral, which can be readily estimated accurately by employing the
Gaussian-Chebyshev Quadrature (GCQ) formula \cite[Eq.(25.4.39)]{BibAbramowitzStegunBook}, which converges rapidly and steadily, requiring only few terms for an
accurate result.

%
\begin{figure*}[h]
\begin{tabular}{c}
\begin{minipage}[t]{1\columnwidth}
\begin{center}
\scriptsize
\vskip -1mm
\begin{multline}\label{Eq:EqASEPOverRaylayFadingChannels}
P_{ASEP,AP-1}={-{2}^{N}
\int_{0}^{\infty}Z\left(u\right)\Biggl(
	\BesselK[1]{2\sqrt{\frac{u}{\Omega_{\mathcal{D}}}}}
	\prod_{k=1}^{N-1}\sum_{\ell=1}^{L}(-1)^{\ell+1}\Binomial{L}{\ell}\sqrt{\frac{{\ell}{u}}{\Omega_{k}}}\BesselK[1]{2\sqrt{\frac{{\ell}{u}}{\Omega_{k}}}}
	\Biggr)
}
\manualtimesbreak
{
	\left(
		\frac{1}{\Omega_{\mathcal{D}}}\frac{\BesselK[0]{2\sqrt{\frac{u}{\Omega_{\mathcal{D}}}}}}{\BesselK[1]{2\sqrt{\frac{u}{\Omega_{\mathcal{D}}}}}}+
		\frac{1}{\sqrt{\Omega_{\mathcal{D}}}}\sum_{k=1}^{N-1}\frac{
			\sum\limits_{\ell=1}^{L}(-1)^{\ell}{\ell}\Binomial{L}{\ell}\frac{1}{\Omega_{k}}\BesselK[0]{2\sqrt{\frac{{\ell}{u}}{\Omega_{k}}}}
		}{
			\sum\limits_{\ell=1}^{L}(-1)^{\ell}\Binomial{L}{\ell}\sqrt{\frac{{\ell}}{\Omega_{k}}}\BesselK[1]{2\sqrt{\frac{{\ell}{u}}{\Omega_{k}}}}
		}
	\right)
}du.
\end{multline}
\vspace{-1mm}
\end{center}
\end{minipage}
\\
\hline
\end{tabular}
\vskip -32pt 
\end{figure*}

\subsection{AP-2 Protocol}

For the AP-2 protocol, one has

\begin{equation}\label{Eq:EqEndSNRInReciprocalityAP2}
\widetilde{\gamma}_{end}=\sum_{k=1}^{N-2}\widetilde{\gamma}_{k}+\widetilde{\gamma}_\mathcal{P},
\end{equation}
where $\widetilde{\gamma}_\mathcal{P}=\frac{1}{{\gamma}_\mathcal{P}}=\left( \max_{\ell} \{\frac{1}{\widetilde{\gamma}_{N-1,\ell}
+ \widetilde{\gamma}_{N,\ell} } \} \right)^{-1}$. The MGF $\mathcal{M}_{\gammatilde_{end}}(s)$ can be expressed as

\begin{equation}\label{Eq:EqEndToEndMGF_AP2}
\mathcal{M}_{\gammatilde_{end}}(s)=\mathcal{M}_{\mathcal{O}}(s)\times\mathcal{M}_{\mathcal{P}}(s),
\end{equation}
where $\mathcal{M}_{\mathcal{O}}(s)=\prod_{m=1}^{N-2}\mathcal{M}_{m}(u)$ is the MGF of the first term of \eqref{Eq:EqEndSNRInReciprocalityAP2} and
$\mathcal{M}_{\mathcal{P}}$ denotes the MGF of $\widetilde{\gamma}_\mathcal{P}$. From \eqref{Eq:EqHopMGF}, \eqref{Eq:EqHopCDF} and \eqref{Eq:EqMGFOfOneHop}, one
has

\begin{equation}\label{Eq:EqMoMGF}
\mathcal{M}_{\mathcal{O}}\left(s\right)={2}^{N-2}\prod_{k=1}^{N-2}\sum_{\ell=1}^{L}{(-1)^{\ell+1}}\Binomial{L}{\ell}\sqrt{\frac{\ell{s}}{\Omega_{k}}}\BesselK[1]{2\sqrt{\frac{\ell{s}}{\Omega_{k}}}}.
\end{equation}
The MGF of $\widetilde{\gamma}_\mathcal{P}$ is not available in closed form. Thus, to make the analysis tractable, we use the bounds proposed in
\cite{BibKaveh2004} i.e.
\begin{equation}\label{Eq:EqMoMGF}
\frac{1}{2}\min \{{\gamma}_{N-1,\ell} + {\gamma}_{N,\ell} \} \leq \frac{1}{\widetilde{\gamma}_{N-1,\ell} + \widetilde{\gamma}_{N,\ell} } < \min
\{{\gamma}_{N-1,\ell} + {\gamma}_{N,\ell} \}.
\end{equation}

Using the bounds, $\widetilde{\gamma}_\mathcal{P} \approx \widetilde{\gamma}_\mathcal{P}' = \min_{\ell} \{ \frac{1}{\gamma_A(\ell)} \} =  \max_{\ell}
\{{\gamma_A(\ell)} \}$, where ${\gamma}_{A}(\ell) = c \min \{ {\gamma}_{N-1,\ell} + {\gamma}_{N,\ell} \}$ and $c={1}$ in case of the upper bound or
$c=\frac{1}{2}$ in case of the lower bound. It can be easily found that the distribution of $\gamma_A(\ell)$ is exponential with mean
$\mu_{\ell,c}=\frac{c}{\widetilde{\Omega,}_{N-1,{\ell}}+\widetilde{\Omega}_{N,{\ell}}}$. Thus, the MGF of $\widetilde{\gamma}_\mathcal{P}' = \max_{\ell}
\{{\gamma_A(\ell)} \}$ can be expressed, similar to \eqref{Eq:EqMGFOfOneHop}, as 

\begin{equation}\label{Eq:EqMGF_P}
\mathcal{M}_{\mathcal{P}}(s) = 2\sum _{\ell=1}^L (-1)^{\ell+1} \binom{L}{\ell} \sqrt{\frac{\ell s}{\mu_{\ell,c}}}  K_1\left(2 \sqrt{\frac{\ell
s}{\mu_{\ell,c}}}\right).
\end{equation} 
Substituting $\mathcal{M}_{\mathcal{O}}(s)$ and $\mathcal{M}_{\mathcal{P}}(s)$ into \eqref{Eq:EqASEP}, we obtain the bound on ASEP of multihop transmission
using AP-2 protocol and is shown in \eqref{Eq:EqASEP_AP2}\comment{\stepcounter{equation}} at the top of this page. Note that, $c={1}$ in case of the lower bound
on ASEP or $c=\frac{1}{2}$ in case of the upper bound on ASEP.

\begin{figure*}[t]
\begin{tabular}{c}
\begin{minipage}[t]{1\columnwidth}
\begin{center}
\scriptsize
\vskip -1mm
\begin{multline}\label{Eq:EqASEP_AP2}
P_{ASEP,AP-2~Bound}={- \int_{0}^{\infty}Z\left(u\right) \Bigg(
	\left( \sum _{\ell=1}^L  (-1)^\ell \binom{L}{\ell} 2 \frac{\ell u}{\mu_{\ell,c}}  K_0\left(2 \sqrt{\frac{\ell u}{\mu_{\ell,c}}}\right) \right) \prod _{k=1}^{{N}-2}
	\left( \sum _{\ell=1}^L 2 (-1)^{\ell+1} \binom{L}{\ell} \sqrt{\frac{\ell u}{\Omega_k}} K_1\left(2 \sqrt{\frac{\ell u}{\Omega_k}}\right) \right)   
}
+ \\
{	\left.   
\left( \sum _{\ell=1}^L (-1)^{\ell+1} \binom{L}{\ell} 2\sqrt{\frac{\ell u}{\mu_{\ell,c}}}  K_1\left(2 \sqrt{\frac{\ell u}{\mu_{\ell,c}}}\right) \right)
\sum _{i=1}^{N-2} \left (\sum _{\ell=1}^L {2 (-1)^{\ell} \binom{L}{\ell}\frac{\ell}{\Omega_i} K_0\left(2 \sqrt{\frac{\ell u}{\Omega _i}}\right)} \right) 
 \prod_{\substack{k=1 \\ k \neq i} }^{N-2}
 \left( \sum _{\ell=1}^L 2 (-1)^{\ell+1} \binom{L}{\ell} \sqrt{\frac{\ell u}{\Omega_k}} \right) K_1 \left(2 \sqrt{\frac{\ell u}{\Omega_k}} \right)
 \right)
}du.
\end{multline}
\vspace{-1mm}
\end{center}
\end{minipage}
\\
\hline
\end{tabular}
\vskip -12pt 
\end{figure*}

The instantaneous SEP can also be expressed as a special case of the following generic equality
\begin{equation}\label{Eq:EqInstantaneousSEP}
P_{SEP}\left(\gamma_{end}\right)=\sum_{d=1}^{D}\int_{0}^{\theta_{d}}{\alpha_{d}\left(\theta\right)}{\exp{\left(-{\beta_{d}\left(\theta\right)}\gamma_{end}\right)}}{d\theta},
\end{equation}
by utilizing both alternative exponential forms (i.e., Craig formula forms) of the Gaussian Q-function, i.e., \cite[Eq.(4.2)]{BibAlouiniBook}
and \cite[Eq. (4.9)]{BibAlouiniBook}. Furthermore, the parameters $D$, $\theta_d$, $\alpha_d\left(\theta\right)$ and $\beta_d\left(\theta\right)$
\cite{BibAlouiniBook} are the well-known performance parameters for various modulation schemes \cite{BibAlouiniBook} (i.e., M-ary phase shift keying (M-PSK),
M-ary quadrature amplitude (M-QAM), M-ary pulse amplitude (M-PAM), and so on) and they are independent of instantaneous SNR \cite{BibAlouiniBook}. For binary
modulation schemes, \eqref{Eq:EqInstantaneousSEP} simplifies into   
\begin{equation}\label{Eq:EqInstantaneousSEPWithIncompleteGamma}
P_{SEP}\left(\gamma_{end}\right)=\frac{\Gamma\left(b,a\gamma_{end}\right)}{2\Gamma\left(b\right)},
\end{equation}
where $\Gamma\left(\cdot,\cdot\right)$ is the complementary incomplete Gamma function \cite[Eq.(18.48)]{BibZwillingerBook}. The parameters $a$ and $b$ take
specific values for specific modulations. For instance, $a=1$ is for binary PSK (BPSK) and $1/2$ for binary FSK (BFSK), and $b=1$ is for noncoherent BFSK
(NCFSK)/ differentially coherent BPSK (DPSK) and $1/2$ is for coherent BFSK/BPSK. For this instantaneous SEP, the auxiliary function
$\mathcal{Z}\left(u\right)$ can be given by \cite{BibFerkan2010}  
\begin{equation}\label{Eq:EqZuComplementaryIncompleteGammaForm}
\mathcal{Z}\left(u\right)=\frac{1}{2\Gamma\left(b\right)}\MeijerG[right]{2,0}{1,3}{au}{1}{b,0,0}.
\end{equation}  
Thus, by substituting \eqref{Eq:EqZuComplementaryIncompleteGammaForm} in \eqref{Eq:EqASEPOverRaylayFadingChannels} or \eqref{Eq:EqASEP_AP2}, one obtains unified
ASEP expressions for all binary modulation schemes for the AP-1 protocol or the AP-2 protocol, respectively.


\begin{figure}
\begin{center}
\begin{tabular}{l}
\hbox{
\begin{minipage}{0.8\linewidth}
\begin{center}
\includegraphics[width=1\columnwidth, keepaspectratio=true]{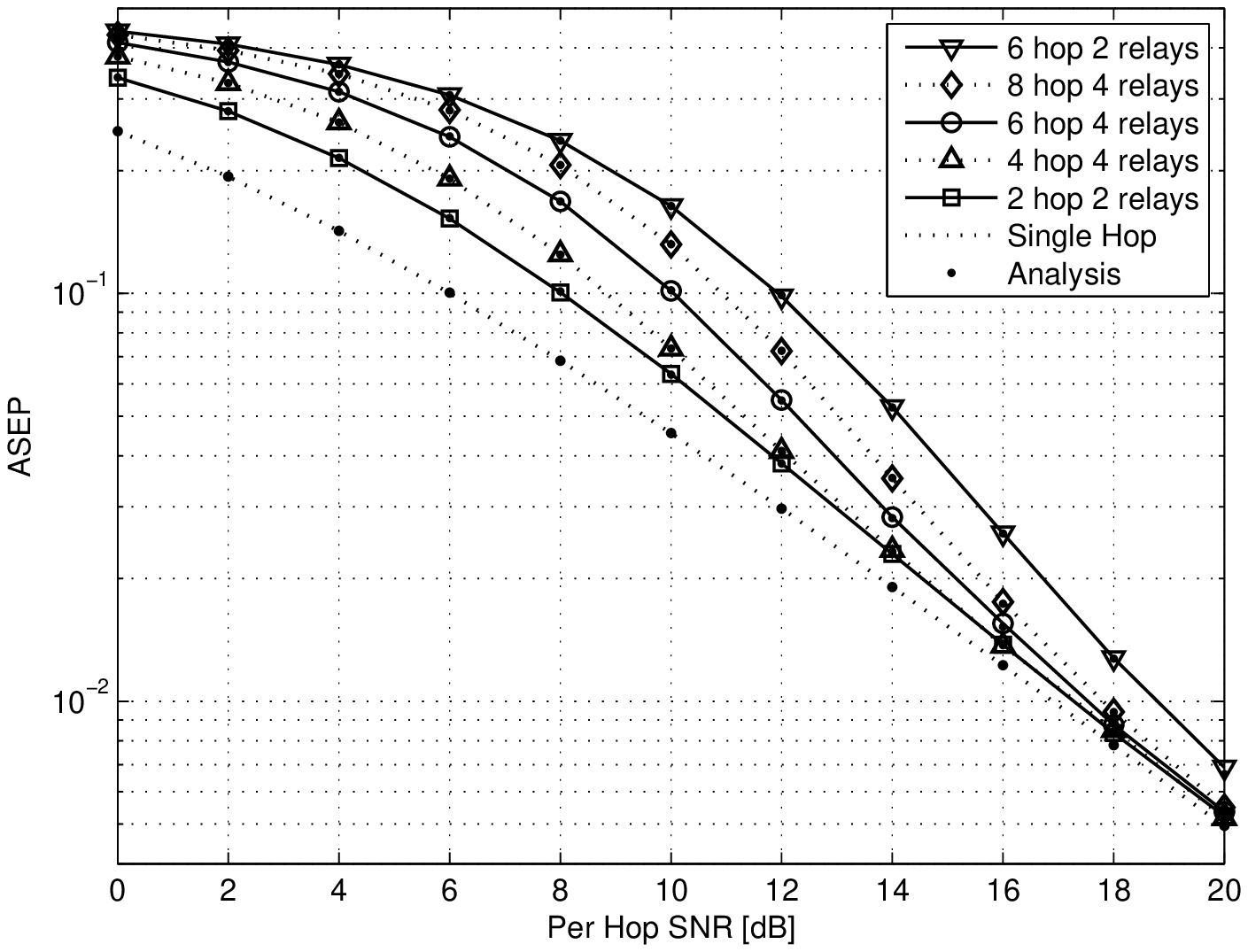}
\caption{ASEP performance comparison of AP-1 protocol.}
\label{Fig:AP1}
\end{center} 
\end{minipage} 
}
\end{tabular}
\end{center}
\vskip -35pt
\end{figure}

\begin{figure}
\begin{center}
\begin{tabular}{l}
\hbox{
\begin{minipage}{0.8\linewidth}
\begin{center}
\includegraphics[width=1\columnwidth, keepaspectratio=true]{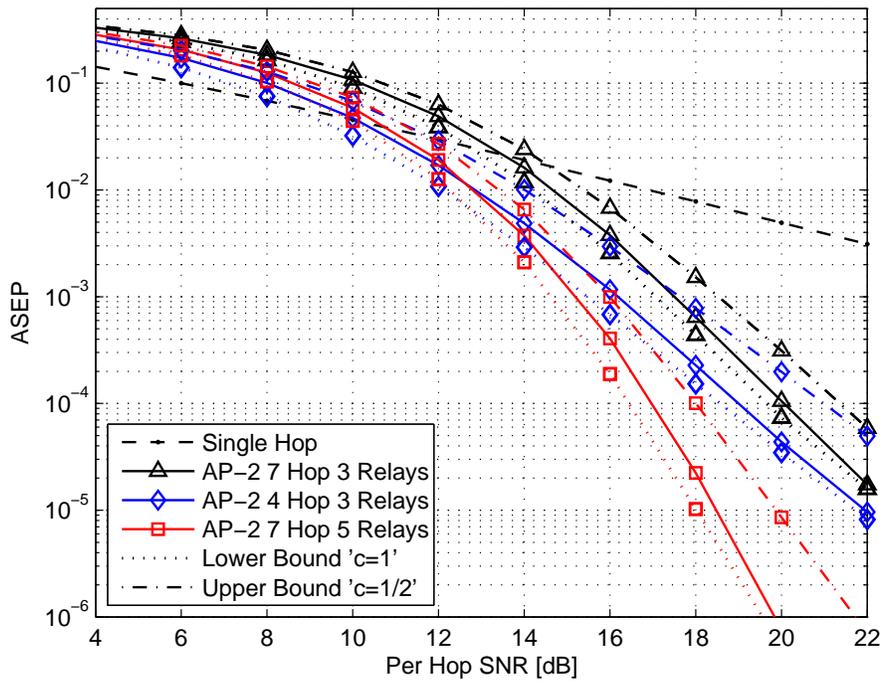} 
\caption{ASEP performance comparison of AP-2 protocol.}
\label{Fig:AP2LB}
\end{center}
\end{minipage} 
}
\end{tabular}
\end{center}
\vskip -35pt
\end{figure}

\subsection*{Numerical Examples}

The performance of the AP-1 protocol and the AP-2 protocol is analysed by Monte-Carlo simulation. The modulation scheme used in the simulations is DPSK.
\figref{Fig:AP1} shows the performance of the AP-1 protocol for varying number of hops and varying number of relays per cluster. It can be observed that the ASEP performance
improves when the number of relays is increased and degrades when the number of hops is increased. Furthermore, the ASEP performance given by
\eqref{Eq:EqASEPOverRaylayFadingChannels} matches exactly with the simulation results. Another important observation is that in all the cases, at high SNR the
ASEP performance converges to the performance of a singlehop system. The ASEP performance is limited because by using the AP-1 protocol there is only one
channel available for transmission at the last hop. Thus, there is no diversity and no performance gain is obtained. In the next section, we propose new
protocols that avoid this problem and give improved performance. The ASEP performance of the AP-2 protocol is shown in \figref{Fig:AP2LB}. The upper bound and
lower bound on the ASEP are also shown. Comparing with \figref{Fig:AP1}, the performance of the AP-2 protocol is much better than the AP-1 protocol. The lower
bound is tight at high SNRs. However, the upper bound tends to be loose, especially at high SNR.

\section{Opportunistic Routing Algorithms}

In this section, new routing protocols are proposed. The routing protocols can be classified into two categories; 1) the unidirectional forward/backward routing
(FR/BR) protocols and 2) the bidirectional forward-backward routing (FBR) protocols. In the unidirectional FR protocol the best path is chosen from the source
to the destination (i.e. a forward scan) where as in the unidirectional BR protocol the best path is chosen from the destination to the source (i.e. a backward
scan). The FR and BR protocols give similar performance if the network is symmetric. In the FBR protocols the FR and BR protocols are applied to the network to
obtain two paths. One of the two paths is then selected for transmission, resulting in a diversity gain and improved performance. We propose the last-$n$-hop
selection protocol, denoted by AP-$n$ and the dual path protocol, denoted by DPP, as FR/BR protocols as well as their FBR variants to improve the performance
of the wireless relay network. 

\subsection{Last-$n$-Hop Selection Protocol}
As was shown in the previous section that the best performance that a multihop clustered system can achieve using the AP-1 protocol is that of a single hop
system. In order to improve the system performance, the adhoc routing protocol was proposed in \cite{BibBoGui2009}. We propose a generalized version of the
adhoc routing protocol in which the best path is chosen from the last $n$ hops instead of only the last 2 hops. Using the AP-$n$ protocol the path through the
last $n$ hops is chosen from $L^{n-1}$ possible paths and a higher diversity gain is achieved as compared with the adhoc protocol proposed in
\cite{BibBoGui2009}. However the performance of the system in this case is limited by the performance of the first $N-n$ hops. For $n=2$, it was shown in
\cite{BibBoGui2009} that a diversity order of $L$ can be achieved at high SNR. If $n$ is increased the performance of the network is improved until for $n=N$
the last-$n$-hop selection protocol becomes exactly the same as the optimal routing protocol \cite{BibBoGui2009}.
\subsubsection{Numerical Examples}
\begin{figure}
\begin{center}
\begin{tabular}{l}
\hbox{
\begin{minipage}{0.8\linewidth}
\begin{center}
\includegraphics[width=1\columnwidth, keepaspectratio=true]{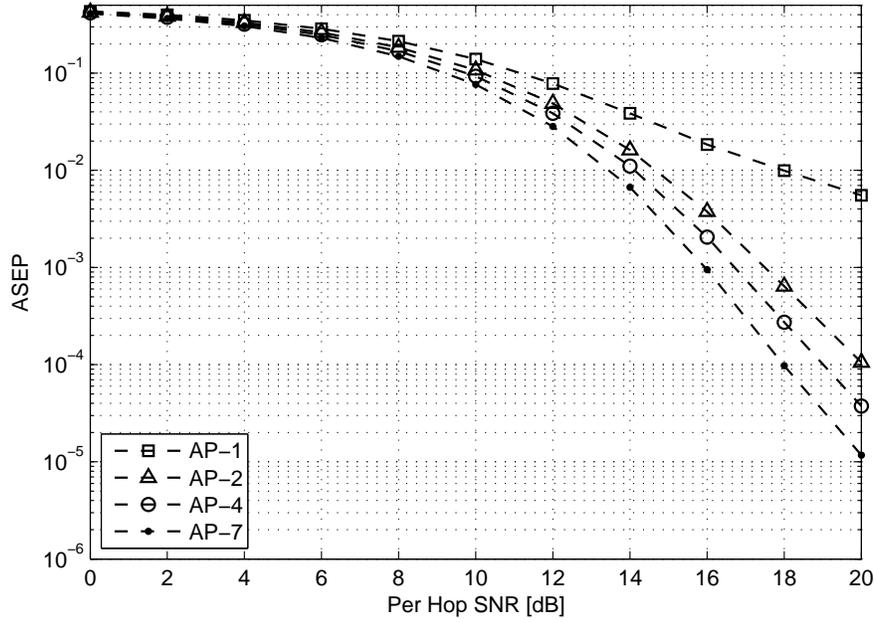}
\caption{ASEP performance comparison of a $7$ hop system with $3$ relays: AP-$n$ protocol with varying values of $n$.}
\label{Fig:Fig002}
\end{center}
\end{minipage}
}
\end{tabular}
\end{center}
\vskip -35pt
\end{figure}

Monte Carlo simulations were carried out to observe the performance of last-$n$-hop selection protocol. The ASEP performance of a 7-hop system with 3 relays per
cluster and DPSK modulation is shown in \figref{Fig:Fig002}. It can be observed that by increasing $n$, the ASEP performance of the system is improved. The
largest gain in performance is observed when $n$ is increased from $1$ to $2$. The relative performance gain is reduced as $n$ is increased from 2. For $n=7$,
the protocol becomes the optimal routing protocol.

\subsubsection{Implementation Issues}
The AP-$n$ protocol can be implemented using a centralized controller or a combination of a centralized controller and a distributed protocol for routing
\cite{BibBoGui2009}. The CSI of all the paths is available to the central controller and it decides the best path to transmit the signal. The base station or an
access point can be used as a central controller \cite{BibBoGui2009}.

Assuming that every transmitting node has knowledge of the CSI of the channels to the relays of the adjacent cluster, the best path for the AP-$1$ system can be
decided in a distributed way i.e. at every stage the transmitting node chooses the path with highest SNR. It broadcasts the a packet to the adjacent cluster,
during a guard time, and the relay for which the packet is meant comes to a listening mode. The node then transmits to the relay in the next cluster. At every
hop a similar procedure is used.

For the distributed implementation of the AP-$2$ system, the same method as described above can be used to find the path through the first $N-2$ hops. However
for the last two hops the distributed protocol proposed in \cite{BibBletsas2006} can be used. A semi-distributed protocol can also be employed in which
a centralized controller is used to find the path through the last $n$ hops. If $n$ is increased from 2, the protocol to select the path in the last $n$ hops
cannot be implemented in a distributed way. Therefore a centralized controller should be used to calculate the path for the last $n$ hops. For the first $N-n$
hops, either the path can be found either in a distributed way or or a central controller can be used.

\subsection{Dual Path Protocol}

In the case of the dual path protocol, two paths are initiated from the source i.e. the channels with the highest and second highest SNRs are chosen for the two
paths respectively. At the remaining hops the channel with the highest SNR is chosen for both the initiated paths. It is possible that the best channels for
both the paths combine at a node. In this case there will be two options; 1) to continue with the best channel or 2) to force both the paths to split by
choosing the best and the second best channels.

In the first case, after the joining node, both the paths will overlap as the channel in both cases is chosen according to the highest SNR. Whereas for the
second case, the path that was initiated with the highest SNR is assigned the best available channel and the other path is assigned the second best available
channel. In the end the path with the best end-to-end SNR is chosen for transmission. The first case is less complex but does not result in much gain in
performance as the paths combine at some node and the performance will be limited by the last hop. However in the second case this is avoided and significant
performance gain can be obtained. Thus, in this paper, we only consider the second case as the dual path protocol and denote it by DPP.

\subsubsection{Numerical Examples}
\begin{figure}
\begin{center}
\begin{tabular}{l}
\hbox{
\begin{minipage}{0.8\linewidth}
\begin{center}
\includegraphics[width=1\columnwidth, keepaspectratio=true]{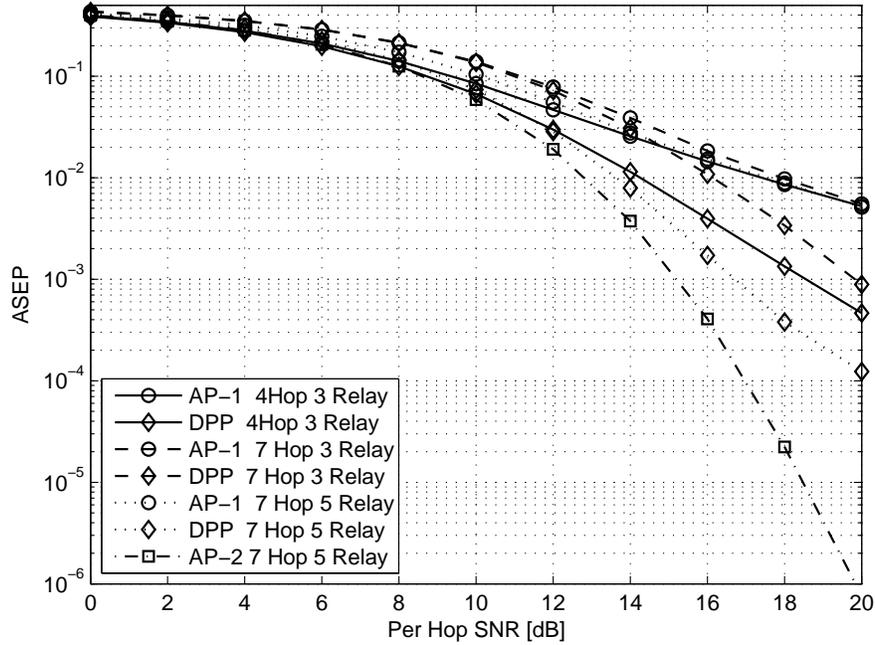}
\caption{ASEP performance of the DPP protocol.}
\label{Fig:Fig003}
\end{center}
\end{minipage}
}
\end{tabular}
\end{center}
\vskip -35pt
\end{figure}

Monte Carlo simulations were carried out to observe the performance of DPP protocol. The ASEP performance is improved by using the DPP protocol as can be
observed in \figref{Fig:Fig003}. The performance of the network using the DPP protocol and the AP-$1$ protocol is compared. Three networks are considered; a
$4$-hop system with $3$ relays per cluster, a $7$-hop system with $3$ relays per cluster and a $7$-hop system with $5$ relays per cluster. All the systems
are assumed to employ DPSK modulation. It can be observed that the performance of the AP-1 protocol is limited for all the three systems and at high SNR they
converge to the same ASEP. The DPP protocol performs better than the AP-$1$ protocol for all the systems and a gain of at least $4$ dB can be observed at
around $20$ dB SNR. It can be observed that at low SNR the 4-hop system performs the best but at high SNR the performance of the 7-hop systems  is better. The
reason for this is that the number of paths that can be obtained through the 7-hop systems is $7^3$ or $7^5$ which is greater than $4^3$ in the case of the
4-hop system and thus there is more diversity. The figure also shows the comparison with the AP-$2$ protocol and it can be observed that the performance of
the DPP protocol is within $2$ dB of the performance of the AP-$2$ protocol at low SNRs.

\subsubsection{Implementation Issues}
The DPP protocol can be implemented using a centralized controller which has the CSI of all the channels in the network. The controller can evaluate the two
paths from the source to destination by using the described DPP protocol and then choose one of the paths based on the higher end-to-end SNR. 

\subsection{Forward-Backward Last-$n$-Hop Selection Protocol} 

A performance gain can be achieved by selection/combining of the paths obtained from the forward and backward scan of the multihop clustered
network. If the paths obtained from the forward and backward scans are completely correlated i.e. exactly the same, no performance gain is achieved. The
maximum diversity gain can be achieved from the FBR strategy provided the forward and backward paths are completely independent.

The signals on the forward and backward paths can be combined using the well known diversity combining techniques such as, $1)$ maximal ratio combining (MRC),
$2)$ selection combining (SC) and $3)$ switching (SSC) \cite{BibAlouiniBook}. The highest performance gain is obtained by using MRC and the least is obtained by using
SSC. The threshold determines the performance of the FBR protocol with SSC. If the threshold is selected too high then there is no switching and almost no diversity gain
is obtained. If the threshold is chosen very low, again no diversity gain is achieved due to continuous switching. We denote this protocol by FBAP-$n$.

\subsubsection{Numerical Examples}
\begin{figure}
\begin{center}
\begin{tabular}{l}
\hbox{
\begin{minipage}{0.8\linewidth}
\begin{center}
\includegraphics[width=1\columnwidth, keepaspectratio=true]{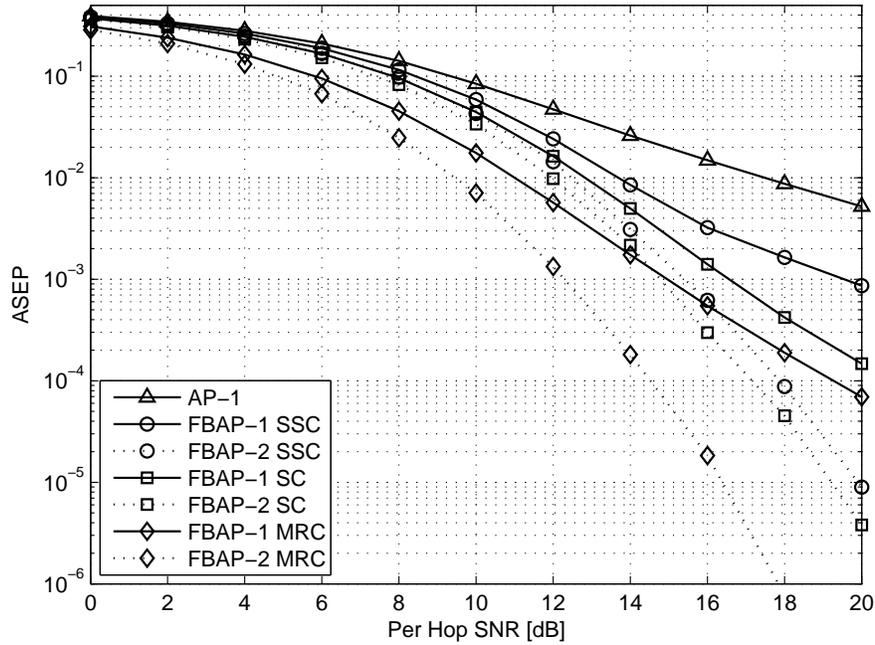}
\caption{ASEP performance comparison of a $4$ hop system with $3$ relays: FBAP-$n$ protocol with varying values of $n$.}
\label{Fig:Fig004}
\end{center}
\end{minipage}
}
\end{tabular}
\end{center}
\vskip -35pt
\end{figure}

Monte Carlo simulations were carried out to observe the performance of proposed FBAP-$n$ protocol. A 4-hop system with 3 relays and employing DPSK modulation
was considered in the simulation. The ASEP performance of the FBAP-$n$ protocol with different combining techniques can be observed in \figref{Fig:Fig004}. It
can be observed that the performance of the FBAP-$n$ protocol is better than that of the AP-1 protocol. The highest performance gain is achieved by FBAP-$n$
protocol with MRC and the least gain is observed in the case of the FBAP-$n$ protocol with SSC. Furthermore increasing the value of $n$ improves the
performance of the systems employing the FBAP-$n$ protocol. The best performance is achieved in the case of the FBAP-$2$ protocol with MRC. Comparing the
performance of the FBAP-1 protocol with the performance of the DPP protocol in \figref{Fig:Fig003} for the 4-hop system with 3 relays, one can observe that
the FBAP-1 protocol performs slightly better than the DPP protocol which is obvious as the FBAP-1 protocol finds the forward and backward paths using the
best available channels whereas the DPP protocol considers the best and the second best channels. 

\subsubsection{Implementation Issues}
The forward and backward paths can be calculated by the central controller to which the CSI information of all the paths is available. The controller can
use any combining technique depending on the system and channel conditions. In the case of selection combining the controller chooses the path with the higher
end-to-end SNR from the forward and the backward paths. For SSC combining the controller compares the end-to-end SNR of the forward path to a certain threshold.
If the SNR is greater than the threshold, the signal is transmitted using the forward path. Otherwise the signal is transmitted via the backward path.

The protocol for data transmission in case of the MRC should be modified slightly as the signal cannot be transmitted using both the paths at the same time. We
propose to divide the full time slot into two equal portions. In the first portion transmission is achieved using the forward path and in the second portion the
data is transmitted using the backward path. This way signal can be combined using MRC at the destination after one complete time slot. The drawback of this
scheme is that the throughput of the system is halved.

\subsection{Forward-Backward Dual Path Protocol} 

The FBR strategy is applied to the DPP protocol to improve the performance of the wireless network. We denote this protocol by FBDPP. The signals on the
forward and backward paths can be combined using the well known diversity combining techniques. Similar to the FBAP-$n$ protocol, MRC will result in the
highest performance gain where as switching gives the least performance gain.

\subsubsection{Numerical Examples}
Monte Carlo simulations were carried out to observe the performance of proposed FBDPP protocol as shown in \figref{Fig:Fig006}. The combining strategy
employed in the simulation was selection combining. Three network configurations each using DPSK modulation are considered; a 4-hop system with 5 relays per
cluster, a 7-hop system with 3 relays per cluster and a 7-hop system with 5 relays per cluster. The performance is compared with the performance of the
FBAP-$1$ protocol. At low SNRs the largest performance gain is observed in the case of the 4-hop system with 5 relays where as the performance gain
is minimal in the case of the 7-hop system with 3 relays. However at higher SNRs the performance gain of the 7-hop system with 5 relays is the largest, as
the number of possible paths is the largest for this network.

\begin{figure}[h]
\begin{center}
\begin{tabular}{l}
\hbox{
\begin{minipage}{0.8\linewidth}
\begin{center}
\includegraphics[width=1.0\columnwidth, keepaspectratio=true]{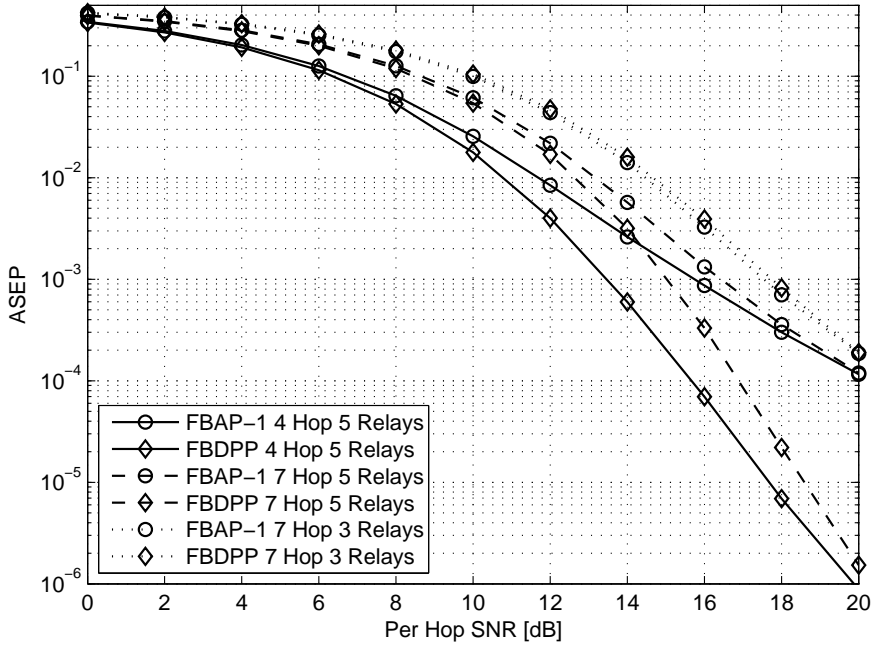}
\caption{ASEP performance of the FBDPP protocol with SC.} 
\label{Fig:Fig005}
\end{center}
\end{minipage}
}
\end{tabular}
\end{center}
\vskip -12pt
\end{figure}

\subsubsection{Implementation Issues}
The best forward and backward paths can be found using the central controller as was suggested for the case of FBAP-$n$ protocol. For selection combining the
controller chooses the best path from the forward and the backward paths. For the switching system it compares the path SNR to a certain threshold and one
of the two paths is chosen accordingly. For MRC, the protocol for data transmission has to be modified as was done for the FBAP-$n$ protocol.


\section{Complexity Analysis}
The complexity of all studied protocols is compared in terms of a) number of CSI required and b) the number of comparison required in finding the best path
through the network. The complexity for all the protocols is briefed in Table 1.

The complexity of the AP-$n$ protocol varies quadratically with $L$ (if $n>2$) and linearly with modifying $N$. The number of CSI required is
$L(N-n)+2L+(n-2)L^2$ and the number of comparisons required is $(L-1)(N-n)+(2L-1)(L(n-2)+1)$. As a special case the complexity of the AP-$1$ and the AP-$2$
protocols is mentioned in Table $1$. For $n=N$ the complexity of the AP-$n$ protocol is the same as the complexity of the optimal protocol.  The complexity
of the FBAP-$n$ protocol is approximately twice the complexity of the AP-$n$ protocol and it can get twice as great as the optimal system when $n=N$. The CSI
information required is $2L^2 (n-2)+(2L-1)(N-2n+3)+1$  and the number of comparisons required is $2((L-1)(N-n)+(2L-1)(L(n-2)+1))$. For $n=2$ the complexity of
this protocol is linear. As $n$ is increased the complexity increases quadratically with $L$.

The complexity of the DPP and the FBDPP protocols is probabilistic as the combining of the paths is random. Assuming that all the channels in one hop are
independent and identically distributed, the probability that the channels combine at the same node is $p_c = \frac{1}{L^2}$. The number of CSI required for
the DPP protocol is almost twice that of the AP-1 protocol and is almost equal to that of the FBAP-1 protocol whereas the number of comparisons required is
more than those required in the case of the FBAP-1 protocol and the protocol is thus more complex. The FBDPP protocol is approximately twice more complex
then the DPP protocol and also more complex then the FBAP-$n$ protocol.


%
\begin{table*}[t]
\begin{center} 
	\caption{Complexity comparison of different protocols for transmission over multihop relay network} 
	\resizebox{1.01\textwidth}{!}{
	\begin{tabular}{lll}
		\toprule
		\textbf{\textsc{ Routing Protocols}} & \textbf{\textsc{ Number of CSI}} & \textbf{\textsc{ Number of Comparisons}}\\
		\toprule
		~~Optimal Routing \cite{BibBoGui2009} & ~~$L^2 (N-2)+2L$ & ~~$2L^2 (N-2)-L(N-4)-1$\\
		\midrule
		~~AP-$1$ \cite{BibBoGui2009} & ~~$L(N-1)$ & ~~$(L-1)(N-1)$\\
		\midrule
		~~FBAP-$1$ & ~~$(2L-1)(N-1)+L$ & ~~$2(L-1)(N-1)+1$\\
		\midrule 
		~~AP-$2$ \cite{BibBoGui2009} & ~~$LN$ & ~~$N(L-1)+1$\\
		\midrule
		~~FBAP-$2$ & ~~$LN+(L-1)(N-1)$ & ~~$2(N(L-1)+1)+1$\\
		\midrule
		~~AP-$n$  & ~~$L(N-n)+2L+(n-2)L^2$ & ~~$(L-1)(N-n)+(2L-1)(L(n-2)+1)$\\
		\midrule
		~~FBAP-$n$ & ~~$2L^2 (n-2)+(2L-1)(N-2n+3)+1$ & ~~$2((L-1)(N-n)+(2L-1)(L(n-2)+1))+1$\\
		\midrule
		~~DPP	 & ~~$L(2N -1)+2$ & ~~$ (L+1) +(N-2)(2(L-1)+p_c(L-2))$\\
		\midrule
		~~FBDPP   & ~~$LN + 2((N-1)(L-1)+1)$ & ~~$2((L+1) +(N-2)(2(L-1)+p_c(L-2)))+1$\\
		\bottomrule 
	\end{tabular}}
\end{center}
\vskip -12pt
\end{table*}  
\section{Performance Comparison}

\begin{figure}
\begin{center}
\begin{tabular}{l}
\hbox{
\begin{minipage}{0.8\linewidth}
\begin{center}
\includegraphics[width=1.0\columnwidth, keepaspectratio=true]{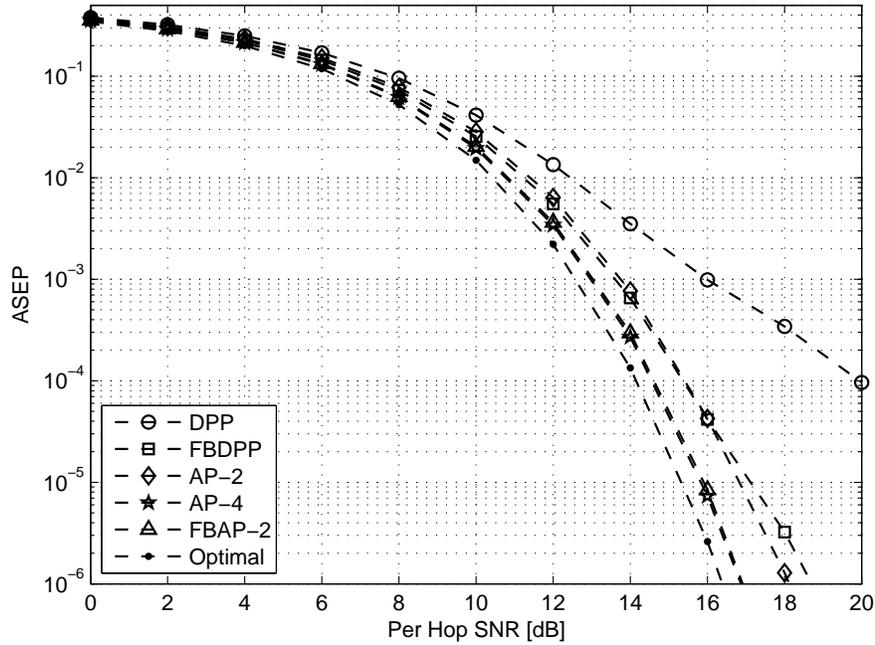}
\caption{ASEP performance comparison of a $6$ hop system with $7$ relays: AP-$n$ vs FBAP-$2$ with SC vs DPP vs FBDPP with SC.} 
\label{Fig:Fig006}
\end{center}
\end{minipage} 
}
\end{tabular}
\end{center}
\vskip -30pt 
\end{figure}

The ASEP performance of the AP-$2$, the AP-$5$, the FBAP-$2$, the DPP, the FBDPP and the optimal routing protocol, for a $6$-hop system with $7$ relays using
DPSK modulation, is shown in \figref{Fig:Fig006}. It can be observed that the performance gain of the DPP protocol is the least as compared with the other
protocols. The FBDPP protocol performance is almost same as the performance of the AP-2 protocol. However the FBDPP protocol has higher complexity. The
ASEP performance of the FBAP-$2$ routing protocol and the AP-$4$ routing protocol is almost the same and is close to the performance of the optimal routing
protocol. The number of CSI required for the AP-4 protocol is 48 which is higher than 36 which is the number of CSI required for the FBAP-2 protocol. The
number of comparisons required for the FBAP-$2$ system are $39$ which is less than the number of comparisons required for the AP-$4$ protocol, i.e. $69$. Thus
the FBAP-$2$ protocol is less complex than the AP-$5$ protocol and gives better ASEP performance. Therefore in order to achieve close to optimal performance
the FBAP-$2$ routing protocol is a better choice than the increasing $n$ in AP-$n$ protocol or employing the FBDPP protocol.

\section{Conclusion}
The average symbol error probability expression for the AP-1 protocol and the AP-2 protocol were derived. These expressions are useful to evaluate the
performance of these protocols. It was shown that routing protocols have large impact on the system performance. Devising better routing protocols can
greatly improve the performance of the multihop relay network. We proposed four routing protocols, namely; 1) the last-$n$-hop selection protocol, 2) the dual
path protocol, 3) the forward-backward last-$n$-hop selection protocol and 4) the forward-backward dual path protocol.  For near optimal performance higher
complexity protocols such as FBAP-$2$ can be used. AP-$2$ protocol or AP-$1$ protocol can be used in situations where the system cannot support high
complexity and a compromise for performance can be made. 

\ifCLASSOPTIONcaptionsoff
\newpage
\fi
\ifCLASSOPTIONcaptionsoff
\newpage
\fi

\bibliography{IEEEfull,document}
\bibliographystyle{IEEEtran}


\end{document}